\documentclass[aps,prl,superscriptaddress,showpacs,reprint]{revtex4-1}
\usepackage{graphicx}
\usepackage{amsmath}
\usepackage{amssymb}

\begin{document}

\title{Comment on ``Coulomb Instabilities of a Three-Dimensional Higher-Order Topological Insulator''}

\author{Yu-Wen Lee}
\affiliation{Department of Applied Physics, Tunghai University, Taichung, Taiwan, Republic of China}

\author{Min-Fong Yang}
\affiliation{Department of Applied Physics, Tunghai University, Taichung, Taiwan, Republic of China}

\date{\today}

\maketitle

Recently it was concluded based on renormalization-group (RG) calculations that three-dimensional second-order topological insulators (SOTIs) are always unstable to the Coulomb interaction~\cite{PLZhao_etal2021}. They will undergo  topological phase transitions to either topological insulators (TIs) or normal insulators (NIs).
In this Comment we point out that 
the SOTIs are actually robust against weak Coulomb interactions and there exists no topological phase transition from SOTIs to TIs within the RG approach.

As shown by Eq.~(3) in~\cite{PLZhao_etal2021}, the low-energy effective action for the SOTI is described by a theory of Dirac fermions with a mass $m$ and with quadratic corrections represented by the $B_i$ and the $D$ terms. For the non-interacting case, 
the nonzero $D$ term 
can induce a sign-changing mass gap for the surface states, leading to in-gap hinge modes at the mass domain walls of surface states~\cite{Schindler2020}. It is shown by Fig.~2(a) of~\cite{PLZhao_etal2021} that $D$ can flow to zero in the low-energy limit in the presence of the Coulomb interaction. This result of vanishing $D$ is used to identify the transition from SOTI to TI.

While calculations in~\cite{PLZhao_etal2021} are correct,
the proposed SOTI-TI transition is not true. This is due to a wrong criterion for this transition being employed.
We note that the renormalized $D$ flows to zero \emph{even for the non-interacting case}, but it does certainly not imply a phase transition to TI ! This result of vanishing $D$ is expected, because the $D$ term describes a quadratic correction to the Dirac theory and then is irrelevant around the non-interacting fixed point. This conclusion can be explicitly drawn as well from RG equations in Supplemental Material of~\cite{PLZhao_etal2021}
by setting the parameter $\alpha$ denoting the Coulomb interaction to zero. Therefore, the irrelevance of the $D$ term simply indicates that \emph{it plays no role in the low-energy physics for bulk properties, instead of showing a phase transition at boundaries}.

\begin{figure}[tp]
\includegraphics[width=0.23\textwidth]{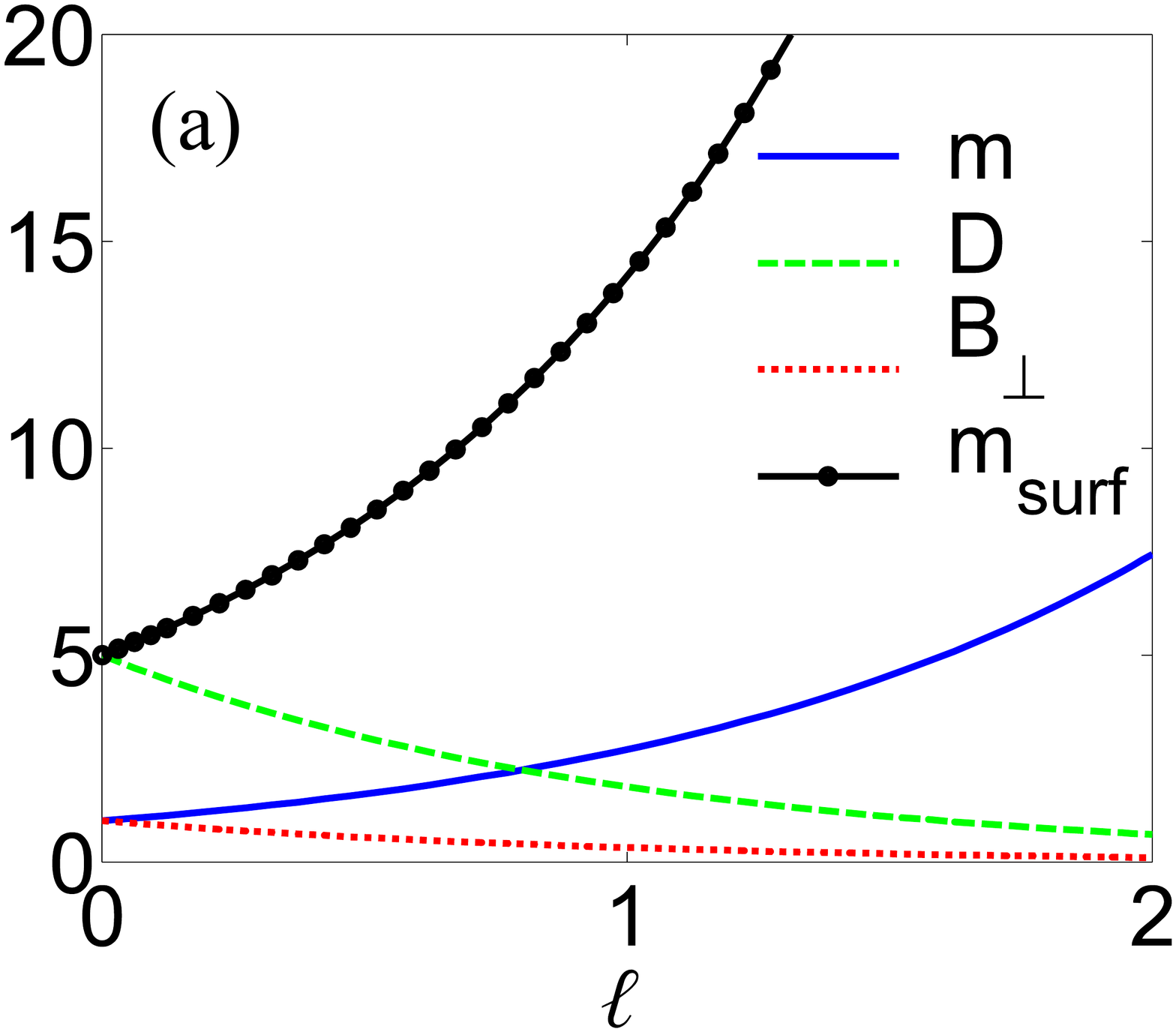}
\includegraphics[width=0.24\textwidth]{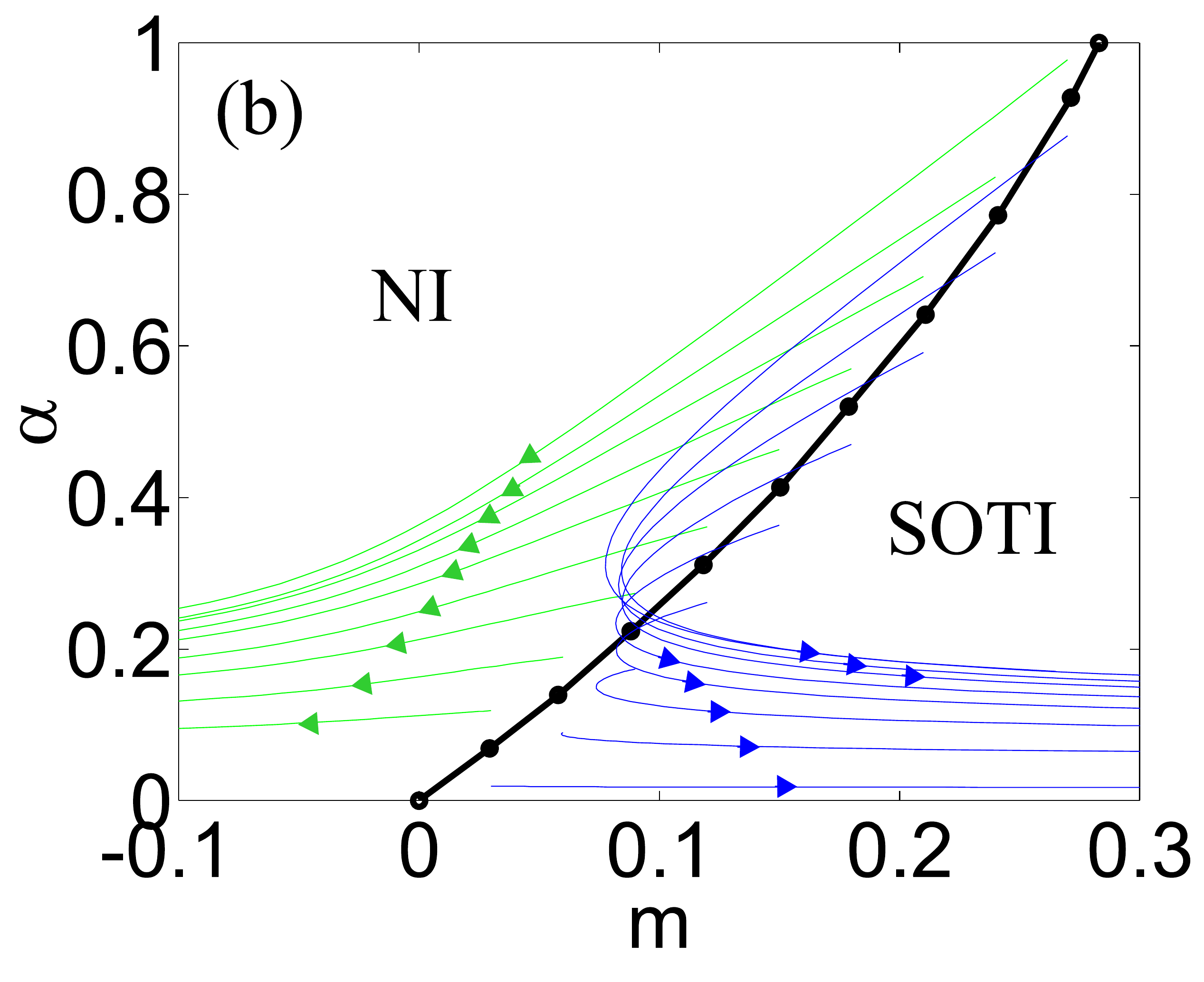}
\caption{%
(a) The renormalized $m$, $D$, $B_\perp$, and $m_\textrm{surf}=mD/B_\perp$ as functions of the running scale parameter $\ell$. 
Here the initial values of system parameters are $m_0=B_{\perp}^{0}=1$, $B_{z}^{0}=0.5$, $\alpha_0=\gamma^2_0=0.1$, and $D_0=5$.
(b) The RG flow (blue and green lines) and the phase boundary (black line) in the $m-\alpha$ plane. Here the same parameters as those in Fig.~4(c) of~\cite{PLZhao_etal2021} are used.
}\label{fig1}
\end{figure}

To have a true SOTI-TI transition, the surface gaps need to be closed, while the mass gap of the bulk states remains finite. For the model under consideration, by using degenerate perturbation theory (for example, see Refs.~\cite{Shen_textbook,Yan_etal2018} and App. A) with the $D$ term being treated as a perturbation, the surface mass can be derived and it becomes $m_\textrm{surf}=mD/B_\perp$ (cf. Eq.~(10) in~\cite{Yan_etal2018}).
Since both $D$ and $B_\perp$ could flow to zero, the vanishing $D$ does not immediately imply the closure of $m_\textrm{surf}$. For illustration, the RG flows of $m$, $D$, $B_\perp$, and $m_\textrm{surf}$ are shown in Fig.~\ref{fig1}(a), where the same initial values of system parameters as those in Fig.~2 of~\cite{PLZhao_etal2021} are employed. We find that, while $D$ and $B_\perp$ approach zero, both the bulk mass $m$ and the surface mass $m_\textrm{surf}$ increase under the RG flow. (For further discussions on perturbative RG approach, see App. B.) Therefore, the claimed SOTI-TI transition does not happen.

By numerically solving the RG equations derived in~\cite{PLZhao_etal2021}, we find that 
the SOTI remains stable up to a critical strength of Coulomb interaction. The phase diagram of the disorder-free model is shown in Fig.~\ref{fig1}(b). Just like the case of TI~\cite{Goswami_etal2011}, the effect of the Coulomb interaction (and nonzero $B_i$) is to cause nonuniversal shift of the phase boundary. The phase boundary between NI and SOTI agrees with that depicted in Fig.~4(c) of~\cite{PLZhao_etal2021}, while the interpretation of the phases separated by the boundary is different. The left-hand side of the phase boundary represents the NI phase, in which the renormalized parameters will flow to the fixed point with negative $m$, while the right-hand side shows the SOTI phase characterized by the growing of $m$ and $m_\textrm{surf}$.

In summary, to explore novel phase transitions out of higher-order topological phases of matter, correct physical quantities protecting the higher-order topological properties need to be measured. Otherwise, misleading conclusions could be drawn. In this Comment we would like to remind that, to locate the possible transitions from SOTI to TI, an appropriate quantity is the sign-changing mass gap $m_\textrm{surf}$ for the surface states. By examining its behavior within RG approach, the stability of SOTI against weak Coulomb interactions is demonstrated. Interestingly, a recent work arrives at the same conclusion as ours and thus provides a further support to our results~\cite{Li_etal2022}.

We acknowledge financial support from the Ministry of Science and Technology of Taiwan under Grant No. MOST 110-2112-M-029-003 and MOST 110-2112-M-029-004.


\newpage
\section{Appendix A: Derivation of the surface gap}

We show the derivation of the effective low-energy surface Hamiltonian and the corresponding surface gap on the $(100)$ plane by using the degenerate perturbation method discussed in Refs.~\cite{Shen,Yan,Li_etal2020}. Besides the continuous models, this approach is applicable for the lattice models as well~\cite{Yan,Konig_etal2008}.

For a semi-infinite plane $x\ge 0$, we can replace $k_x\rightarrow -i\partial_x$ and decompose the Hamiltonian as $H=H_{0}+H_{p}$, in which
\begin{eqnarray}
H_{0}(-i\partial_{x},k_y,k_z)&=&(m+B_\perp\;\partial_{x}^{2})\tau_{z}\sigma_{0} -iv\partial_{x}\tau_{x}\sigma_{x}\;,\nonumber\\
H_{p}(-i\partial_{x},k_y,k_z)&=&vk_{y}\tau_{x}\sigma_{y} +v_{z}k_{z}\tau_{x}\sigma_{z} + D\partial_{x}^{2}\tau_{y}\sigma_{0}\;,
\end{eqnarray}
where the insignificant $k_y^2$ term has been omitted. The purpose of this decomposition is to solve $H_0$ first, and then treat $H_p$ as a perturbation, which is justified when $k_y$, $k_z$, and $D$ are relatively small.

We expect surface states to be exponentially localized on the boundary, so we look for solutions with the ansatz, $\psi(x)\propto e^{-\lambda x}$. Solving the eigenvalue equation $H_{0}\psi(x)=0$ for the expected surface states with \emph{zero energy}~\cite{note}, we have two solutions under the boundary condition $\psi(0)=\psi(+\infty)=0$,
\begin{eqnarray}
\psi_{\pm}(x)=\phi(x)\;\chi_{\pm} \;,
\end{eqnarray}
where the eigenvectors $\chi_{\alpha=\pm}$ are
\begin{eqnarray} \chi_{+}=|\tau_{y}=+1\rangle\otimes|\sigma_{x}=+1\rangle\;,\nonumber\\
\chi_{-}=|\tau_{y}=-1\rangle\otimes|\sigma_{x}=-1\rangle \;,\nonumber
\end{eqnarray}
satisfying $\tau_{y}\sigma_{x}\chi_{\alpha}=+\chi_{\alpha}$. The spatial wavefunction is $\phi(x)=\mathcal{N} (e^{-\lambda_{1}x}-e^{-\lambda_{2}x})$, which satisfies the differential equation $[(m+B_\perp\;\partial_{x}^{2}) +v\partial_{x}]\phi(x)=0$ with $\mathcal{N}$ being the normalization constant and
\begin{eqnarray}
\lambda_{1,2}=\frac{v\pm\sqrt{v^2-4mB_\perp}}{2B_\perp}\;.
\label{lambda}
\end{eqnarray}

In this basis $\{\psi_{+}(x), \psi_{-}(x)\}$, the matrix elements of the perturbation $H_p$ are
\begin{eqnarray}
H_{{\rm eff},\alpha\beta}(k_{y},k_{z})=\int_{0}^{+\infty} dx\; \psi^*_{\alpha}(x)\;H_{p}(-i\partial_{x},k_{y},k_{z})\;\psi_{\beta}(x)\;,\nonumber
\end{eqnarray}
therefore, the final form of the effective Hamiltonian is
\begin{eqnarray}
H_{\rm eff}(k_{y},k_{z})= vk_{y}s_{y} - v_{z}k_{z}s_{x} - m_{\rm surf}s_{z}\;,
\end{eqnarray}
where $s_{x,y,z}$ are the Pauli matrices in the basis $\{\psi_{+}(x), \psi_{-}(x)\}$. The surface mass for this effective two-dimensional Dirac theory is
\begin{eqnarray}
m_{\rm surf}=-D\int_{0}^{+\infty}dx \phi(x)^{*}\partial_{x}^{2}\phi(x)
=D\frac{m}{B_\perp}.
\label{mass}
\end{eqnarray}

In the longer version of the Reply to our Comment, the authors give exact solutions of the surface states by using a more complicated approach discussed in Ref.~\cite{Shan_etal2010}. While their expression of the surface mass, $m'_\textrm{surf}=mD/\sqrt{D^2+B_\perp^2}$, is somewhat different to ours, up to leading order in $D/B_\perp$, their result reduces to ours. That is, $m'_\textrm{surf}$ and $m_\textrm{surf}$ differ merely quantitatively, rather than qualitatively. Indeed, under the same RG prescription, both expressions of surface mass show similar behaviors.

\section{Appendix B: further discussions on perturbative RG approach}

We would like to emphasize that the superficial divergence of $m_\textrm{surf}$ (and $m'_\textrm{surf}$ as well) in the $\ell\to\infty$ limit arises from incorrectly extending the perturbative RG results beyond the regime of their applicability. Since the RG equations are derived from a continuous model, they are valid only for $m(\ell)\ll \Lambda$, where $\Lambda\sim 1/a$ is the cut-off of the continuous model.  Beyond the scale $\ell>\ell^*$, where $\ell^*$ is defined by $m(\ell^*)\sim \Lambda$, the RG results can no longer be used and the RG flow should stop running~\cite{noteRG_1,noteRG_2}.
In the Wilsonian point of view, it means that, when one lowers the cut-off to the scale given by the mass parameter, most of important infrared quantum fluctuations have been integrated out and thus the running scale should be fixed at $\ell^*$~\cite{noteRG_2}. The renormalized $B_\perp$ and $D$ (and then the surface mass) at this finite RG scale remain nonzero. This shows that the SOTI is actually stable against weak Coulomb interactions, instead of being unstable toward to TI as concluded in the original PRL article.


\end{document}